\documentclass[lettersize]{IEEEtran}
\usepackage{amsmath,amssymb,amsfonts}
\usepackage{array,booktabs,multirow,tabularx}
\usepackage{algorithm,algorithmic}
\usepackage{graphicx}
\usepackage[breakable]{tcolorbox}
\usepackage{url}
\usepackage{cite}
\usepackage[T1]{fontenc}
\usepackage{newtxmath}
\usepackage{xcolor}
\usepackage[hidelinks]{hyperref}

\hypersetup{
  pdftitle={OrbitWeaver: LLM Agent Planning with Simulation Feedback for Optically Interconnected Orbital GPU Clusters},
  pdfauthor={Pengcheng Luo and Bowen Zhang}
}

\newtcolorbox{prompttemplate}[1]{
  breakable,
  sharp corners,
  colback=black!2,
  colframe=black!40,
  colbacktitle=black!10,
  coltitle=black,
  fonttitle=\bfseries\footnotesize,
  title={#1},
  boxrule=0.4pt,
  left=6pt,right=6pt,top=5pt,bottom=5pt,
  before skip=6pt,after skip=6pt
}

\begin{document}
\title{OrbitWeaver: LLM Agent Planning with Simulation Feedback for Optically Interconnected Orbital GPU Clusters}
\author{Pengcheng Luo and Bowen Zhang%
\thanks{This work was supported by the National Science and Technology Major Project of China under Grant 2025ZD1602400. \emph{(Corresponding author: Bowen Zhang.)}}%
\thanks{Pengcheng Luo and Bowen Zhang are with Ningbo Artificial Intelligence Institute, Shanghai Jiao Tong University, Ningbo 315000, China, and also with the School of Automation and Intelligent Sensing, Shanghai Jiao Tong University, Shanghai 200240, China, and the Key Laboratory of System Control and Information Processing, Ministry of Education of China, Shanghai 200240, China (e-mail: luopeng69131@sjtu.edu.cn; bwz96sco@sjtu.edu.cn).}%
\thanks{The source code will be made available at \protect\url{https://github.com/luopeng69131/OrbitWeaver}.}%
}
\maketitle
\pagestyle{plain}
\thispagestyle{plain}

\begin{abstract}
Space-based data-center designs increasingly consider GPU computing formations connected by inter-satellite optical links. With limited neighboring connections, jobs assigned disjoint GPUs can still contend on shared relay paths. Rank placement and communication priorities jointly affect this contention and the readiness of downstream computation. We present OrbitWeaver, an offline large language model (LLM) agent planner that jointly optimizes rank placement and communication order. Guided by communication structure, execution dependencies, and simulation feedback, the agent constructs and refines plans that coordinate shared-link access with computation progress. This coordination reduces contention and dependency blocking, shortening mean job completion time (JCT). Simulation experiments on optically interconnected orbital GPU clusters show that OrbitWeaver reduces aggregate mean JCT by 6.2--33.5\% relative to the state-of-the-art baselines CASSINI, Crux, and Symphony. These results demonstrate the value of LLM-guided planning for efficient concurrent computing in orbital GPU clusters.
\end{abstract}


\section{Introduction}
\label{sec:introduction}
\subsection{Background and Motivation}
\begin{figure*}[!t]
\centering
\includegraphics[width=\textwidth]{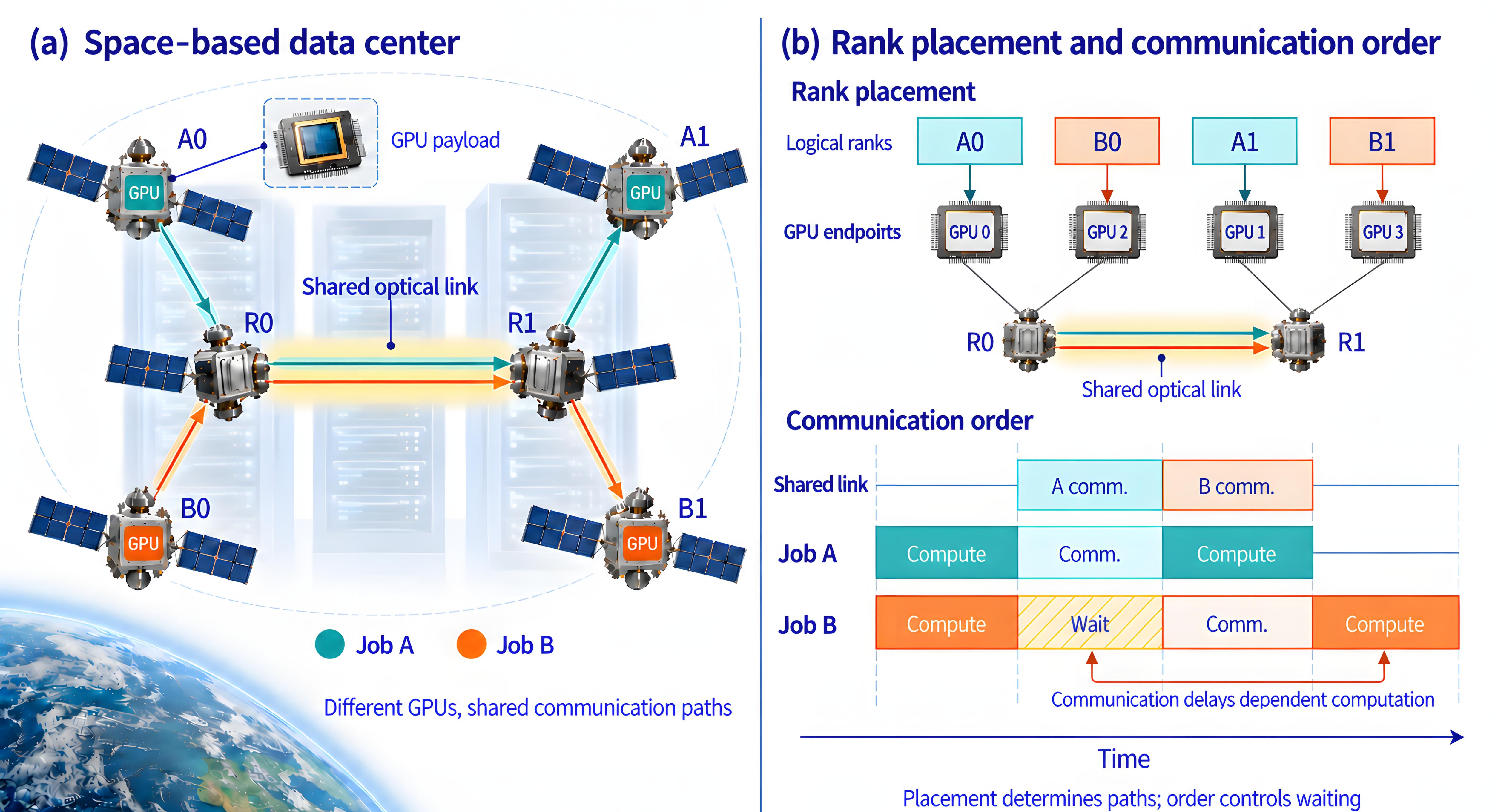}
\caption{Optically interconnected orbital GPU clusters and contention on shared paths.}
\label{fig:background}
\end{figure*}

\IEEEPARstart{T}{he} growing computational and energy demands of artificial intelligence (AI) have motivated interest in space-based data centers. One proposed architecture distributes accelerators across satellites flying in close formation and connects them through optical inter-satellite links, as explored in Google's SunCatcher study~\cite{aguera2025suncatcher}. Fig.~\ref{fig:background}(a) illustrates such an orbital GPU cluster. A distributed job comprises logical processes, or \emph{ranks}, that run on different GPUs, execute computation, and exchange data through collective communication~\cite{wilkins2026acclaim}. When multiple jobs run concurrently, these exchanges share the optical interconnect even if the jobs have separate GPU allocations.

Payload and power budgets, together with formation design, constrain the number and capacity of optical connections~\cite{aguera2025suncatcher}. When each satellite has a limited number of neighboring connections, communication between remote nodes traverses relay paths. In Fig.~\ref{fig:background}(b), jobs assigned disjoint GPUs still contend for shared links, and the resulting waits can delay downstream computation and job completion. Placement affects communication paths and costs~\cite{zhang2023tag}, while communication order affects computation--communication overlap~\cite{duan2023mercury,gao2024usbyte}. These decisions are coupled: remapping ranks can reduce one group's path overlap while creating another, and prioritizing a short collective may not advance the dependencies that determine job completion. Efficient execution therefore requires joint planning of rank placement and communication order, accounting for workload structure, job arrivals, and link capacity.

Job placement, communication scheduling, and workload-oriented interconnect design offer ways to coordinate distributed GPU execution~\cite{rajasekaran2024cassini,liu2024multijob,zhao2025directconnect,peng2019bytescheduler}. Their execution outcomes depend on when communications become ready and how resource waits propagate through computation dependencies, as well as on traffic volume and path length~\cite{hu2022dpro,duan2024proteus}. Static traffic summaries can identify heavily used links, yet they do not by themselves indicate which placement and communication order will minimize job completion time. Joint planning thus calls for a process that connects workload analysis to candidate construction and uses complete-execution feedback to decide which changes to explore next.

Large language model (LLM) agents combine reasoning with tool use and can adapt subsequent actions to returned observations~\cite{yao2023react}. Studies of LLM-driven optimization and GPU kernel generation further demonstrate how candidate scores and execution diagnostics can guide iterative revision~\cite{yang2024opro,ouyang2025kernelbench}. These capabilities suggest a way to address joint planning in orbital GPU clusters: an agent could examine job dependencies and path conflicts, form hypotheses about placement and communication order, and use simulation feedback to select or revise candidates. Execution-graph analysis and complete-job simulation provide observations for this process~\cite{hu2022dpro,won2023astrasim2}. With job graphs available in advance, these interactions can take place offline to produce a validated placement and scheduling plan for subsequent execution.

\begin{figure}[!t]
\centering
\includegraphics[width=\columnwidth]{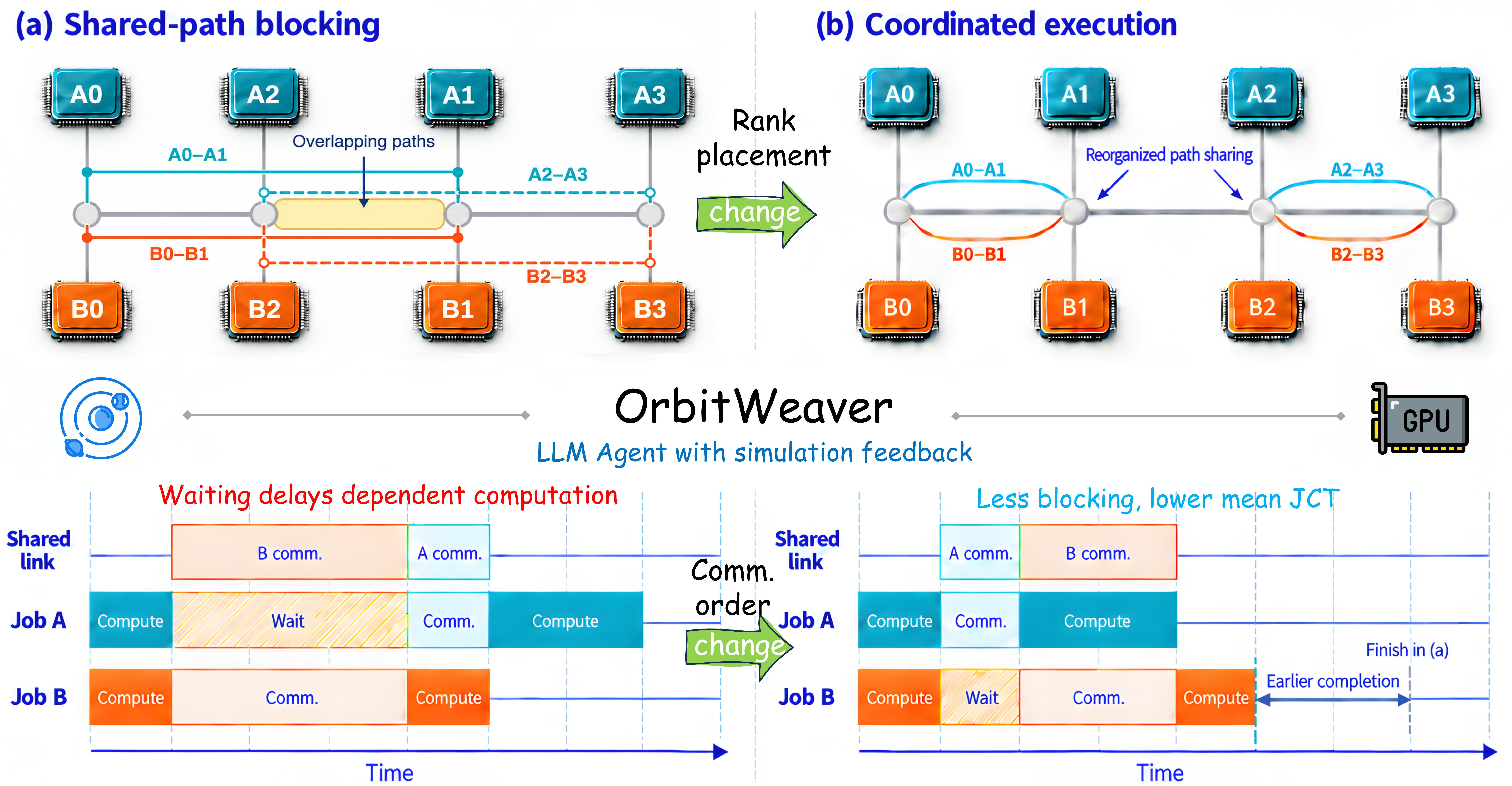}
\caption{Illustrative effects of jointly adjusting rank placement and communication order with OrbitWeaver.}
\label{fig:concept}
\end{figure}

To support this form of joint planning, we present \emph{OrbitWeaver}, an offline LLM agent planner that uses simulation feedback to jointly optimize rank placement and communication order for optically interconnected orbital GPU clusters. It operates on job execution graphs available in advance under fixed GPU allocations and interconnect conditions. Structured tools expose communication membership, path overlap, and execution dependencies to guide candidate construction. Fast simulation supports candidate exploration, while reference simulation provides completion times and diagnostics for revision and validation. Planning concludes with a reference-validated joint plan. Fig.~\ref{fig:concept} illustrates the intended effect: coordinating rank placement and communication order can reduce blocking on shared paths and allow dependent computation to begin earlier.

\subsection{Contributions}
Our main contributions are as follows.
\begin{itemize}
\item We formulate an execution model for orbital GPU jobs on optical interconnects with limited neighboring connections. The model connects rank mapping, shared relay paths, communication admission, and computational dependencies, yielding a joint planning problem for within-job placement and cross-job communication order that minimizes mean job completion time (JCT).
\item We develop OrbitWeaver, an LLM agent planner driven by simulation feedback. Structured tools expose communication subgroups, path conflicts, and dependency waits. The agent uses these observations to construct placement--priority combinations and uses candidate execution results to guide subsequent exploration, producing executable joint plans tailored to each workload.
\item We evaluate OrbitWeaver using three measured GPU execution traces over 16--64 GPUs and four operating conditions. OrbitWeaver achieves the lowest aggregate mean JCT for all three workloads. Ablations further support jointly planning rank placement and communication order to improve concurrent execution.
\end{itemize}

\subsection{Organization}
Section~\ref{sec:related} reviews GPU job planning under interconnect constraints and LLM-driven optimization with execution feedback. Section~\ref{sec:model} presents the orbital GPU cluster and job execution model and formulates the joint planning objective. Section~\ref{sec:method} describes OrbitWeaver's framework, structural analysis, candidate construction, simulation feedback, and plan delivery. Section~\ref{sec:evaluation} presents the experimental setup, evaluates planning effectiveness through overall performance, concurrency levels, operating conditions, bandwidth sensitivity, and ablations, and analyzes planning and simulation overhead. Section~\ref{sec:conclusion} concludes the paper.

\section{Related Work}
\label{sec:related}
Relevant research follows two main lines: GPU job planning under interconnect constraints and LLM-driven optimization with execution feedback. The former studies the relationship between communication structure, resource contention, and job progress; the latter provides mechanisms for candidate construction, tool interaction, and execution evaluation.

\subsection{GPU Job Planning Under Interconnect Constraints}
Spacecraft resource constraints shape both computing performance and cooperation across orbital nodes. Orbital Edge Computing incorporates energy, communication, and orbital constraints into nanosatellite constellation designs for observation processing, supported by simulation using physical models and runtime services~\cite{denby2020orbital}. SatelliteCOTS complements this system perspective with in-orbit measurements of how temperature and energy relate to COTS computing performance~\cite{xing2024satellitecots}. For cooperation across accelerators, SunCatcher explores satellites flying in close formation with inter-satellite optical links~\cite{aguera2025suncatcher}. In such formations, workload planning must account for the optical connectivity available between computing nodes.

Matching communication to this connectivity is also a central concern in terrestrial cluster research. SCOTCH addresses process placement by mapping communication graphs onto architecture graphs through recursive partitioning~\cite{pellegrini1996scotch}. Collective algorithm synthesis adapts the exchanges themselves to physical resources: TACCL uses communication sketches to guide synthesis for a target topology~\cite{shah2023taccl}, while HeteCCL synthesizes schedules under heterogeneous link capacities and routing constraints~\cite{hei2026heteccl}. Zhao \emph{et al.} also treat topology as a design variable, jointly constructing direct-connect networks and collective schedules under limited connectivity to meet workload latency and bandwidth requirements~\cite{zhao2025directconnect}. Together, these studies establish how process mapping, collective algorithms, and topology shape communication paths and costs.

Even with a given mapping and collective algorithm, performance depends on when communication runs relative to computation and other jobs. Within a training job, ByteScheduler combines tensor partitioning, transfer reordering, and dependency proxies to improve computation--communication overlap, with runtime parameters tuned through Bayesian optimization~\cite{peng2019bytescheduler}. Across concurrent jobs, CASSINI exploits periodic communication demand, combining network-aware placement and phase shifts to coordinate access to shared links~\cite{rajasekaran2024cassini}. Crux prioritizes communication according to corrected GPU computation intensity~\cite{cao2024crux}, while Symphony coordinates collectives at runtime using resource availability~\cite{binyahya2025symphony}. These approaches connect communication paths to resource occupancy over time, showing how spatial placement and temporal coordination affect contention and computational progress.

Evaluating this interaction between paths and timing requires a view of complete workload execution. dPRO uses global dataflow graphs, execution replay, and critical-path diagnosis to search for computation, communication, and memory optimizations~\cite{hu2022dpro}. For architecture and performance evaluation, ASTRA-sim2.0 models training dependencies, multidimensional networks, and disaggregated memory through a graph-based execution engine~\cite{won2023astrasim2}, while SimAI integrates training frameworks, compute kernels, and collective libraries to support architecture design and performance tuning~\cite{wang2025simai}. These tools relate local resource decisions to execution across the job graph. In orbital formations with limited neighboring optical connections, jobs with disjoint GPU allocations can still share relay paths. Rank mapping within jobs and communication admission order across jobs therefore jointly affect when downstream operations become ready. Planning on a given optical interconnect requires relating results from complete job execution to these joint decisions as workload structure and arrival patterns vary.

\subsection{LLM-Driven Optimization with Execution Feedback}
LLM-driven optimization connects generation to evidence from external interaction and evaluation. ReAct interleaves reasoning, actions, and observations so that returned information can guide subsequent plans~\cite{yao2023react}. OPRO uses previous candidates and their scores as prompt context for further generation~\cite{yang2024opro}. These approaches establish an interaction in which earlier outcomes inform the next decision, providing a basis for iterative candidate construction and evaluation.

When candidates are executable programs or heuristics, this interaction can guide changes to algorithm structure. FunSearch combines a frozen LLM, a program evaluator, and a candidate database to direct program search~\cite{romeraparedes2024funsearch}. EoH represents candidates through both natural-language descriptions and heuristic code~\cite{liu2024eoh}, while ReEvo derives short- and long-term reflections from relative candidate performance to guide further generation~\cite{ye2024reevo}. LACE organizes heuristic construction through validated interfaces defined by input--output specifications, a tool library, and heuristic composition, evolving complementary heuristics within a runtime budget~\cite{gong2026lace}. Together, these designs connect candidate representations, evaluation feedback, and interface constraints to the algorithmic changes available during search.

Applying this process to system performance requires feedback tied to the execution decisions being changed. KernelBench evaluates generated GPU kernels for functional correctness and performance, using compilation errors, execution results, and profiling information to guide code revisions~\cite{ouyang2025kernelbench}. For scheduling tool operations, Lu and Reda propose LLM-assisted CPU--GPU scheduling with runtime monitoring, selecting CPU execution, immediate GPU execution, or queued GPU execution according to device contention and GPU memory constraints~\cite{lu2026agentic}. The information used for revision thus depends on whether the adjustable decision concerns kernel implementation or resource assignment.

For concurrent jobs with fixed GPU allocations and shared communication paths, the adjustable decisions concern rank placement and communication priority. Relating overall completion scores to communication subgroups, path occupancy, and dependency waits can guide joint changes that account for both contention and downstream progress. Detailed execution evaluation also consumes the budget available for exploring candidates, creating a need to combine inexpensive trials with validation that preserves finer execution semantics. Planning in this setting therefore calls for a connection between structural diagnosis, joint candidate construction, and execution validation within a finite budget.

\section{Orbital GPU Cluster Model and Joint Planning Objective}
\label{sec:model}
This section models job execution in an orbital GPU formation interconnected through a limited number of optical terminals. It describes how rank placement changes relay paths and how communication scheduling affects dependency progress and job completion. We then formulate offline joint planning to minimize mean JCT.

\subsection{Physical Interconnect and Job Execution Graphs}
\label{sec:physical-model}
The physical interconnect determines the resources available for communication across nodes, while execution graphs specify the order of computation and communication. Joint planning connects the two to identify where contention occurs and how waits propagate through dependencies.

We represent an orbital GPU formation with neighboring optical connections as a directed graph $\mathcal H=(\mathcal S,\mathcal L)$. In the evaluated configurations, each node hosts one GPU endpoint and relays traffic along communication paths. The two transmission directions are modeled separately; each directed link $\ell$ has capacity $B_\ell$ and delay $\delta_\ell$.

Neighboring connections are determined by optical terminal configuration and physical design. A planning instance takes established connections and their parameters as input. Topology, capacities, delays, and endpoint routes $P(x,y)$ remain fixed throughout plan execution. Evaluation assumes that power and cooling sustain the nominal computing performance in the recorded traces.

The single-hop delay comprises propagation and a fixed terminal delay:
\[
\delta_\ell=d_\ell/c_0+\tau_\ell^{\mathrm{term}},
\]
where $d_\ell$ is the configured endpoint separation, $c_0$ is the speed of light, and $\tau_\ell^{\mathrm{term}}$ is the fixed terminal delay. This relation connects the interconnect geometry to the communication delay used by the execution model. Section~\ref{sec:workloads} specifies the evaluated parameters.

Let $\mathcal J$ be the set of jobs and $J=|\mathcal J|$. Job $j$ arrives at $a_j$, has logical rank set $\mathcal R_j$, and is assigned a physical GPU set $A_j\subseteq\mathcal S$. Allocations satisfy $|A_j|=|\mathcal R_j|$ and are disjoint across jobs. Each rank executes local operations and exchanges data with other ranks of the same job through collectives. Routes belonging to different jobs can traverse the same relay links, so disjoint GPU allocations still permit cross-job communication contention.

An execution trace defines the job's directed acyclic graph (DAG) $\mathcal G_j=(\mathcal V_j,\mathcal E_j)$. Nodes represent CPU operations, GPU computation, communication, or synchronization. An edge $(u,v)\in\mathcal E_j$ requires $v$ to wait for $u$ to complete. Noncommunication durations $\tau_v$ are given by the trace; synchronization events have zero duration. Each rank has separate execution channels for CPU operations, GPU computation, and GPU communication. Operations execute serially within each channel and may overlap across channels when dependencies permit.

Let $\mathcal C$ be the set of collective operations. The trace specifies each operation's type, data size $m_c$, and participating rank set $M_c\subseteq\mathcal R_{j(c)}$, where $j(c)$ denotes its job. Operation $c$ corresponds to an execution node $v_{c,r}$ on each participating rank $r$. The collective becomes ready only when every participating node satisfies its dependencies. The execution graph thus preserves both local ordering and synchronization across ranks.

\subsection{Rank Placement and Communication Paths}
\label{sec:placement-model}
In an inter-satellite network with limited neighboring connections, the physical locations of communicating ranks determine the relay links they use. Even with fixed GPU allocations, rank mapping changes communication cost and path overlap with other jobs. We therefore connect logical collective membership to physical path demand.

For job $j$, a bijection $\pi_j$ maps ranks to assigned GPUs:
\begin{equation}
\begin{gathered}
\pi_j:\mathcal R_j\longrightarrow A_j,\qquad \pi_j\text{ is bijective},\\
\Pi=(\pi_j)_{j\in\mathcal J}.
\end{gathered}
\label{eq:mapping}
\end{equation}
Each rank occupies one assigned GPU, and two ranks of the same job cannot occupy the same endpoint. Placement optimization retains the assigned GPU sets and changes only rank correspondence within each set.

For collective $c$, a given collective algorithm determines the logical data exchanges. Let $\mathcal T_c$ be the set of directed rank pairs that exchange data under this algorithm. The endpoints and required links are
\begin{equation}
\begin{aligned}
\mathcal P_c(\Pi)&=\{\pi_{j(c)}(r):r\in M_c\},\\
\mathcal L_c(\Pi)&=\bigcup_{(r,r')\in\mathcal T_c}
P\bigl(\pi_{j(c)}(r),\pi_{j(c)}(r')\bigr).
\end{aligned}
\label{eq:paths}
\end{equation}
Overlap between link sets determines which collectives must wait for one another to release resources. Communication duration depends on operation type, data size, the collective algorithm, and path capacities and delays. Placement therefore changes both the operation's execution cost and its resource conflicts with other operations.

\subsection{Dependency Progress and Communication Admission}
\label{sec:admission}
When jobs share relay paths, communication order determines which dependencies are released first. Actual start times depend on predecessor completion and endpoint and link availability. Separating dependency readiness from resource admission captures how communication queues delay subsequent computation and identifies the ordering decisions controlled by priorities.

Let $j(v)$ denote the job containing node $v$, and let $\operatorname{pred}(v)$ be its immediate predecessors. Node dependency readiness and collective all-member readiness are
\begin{equation}
\begin{aligned}
\rho_v&=\max\left\{a_{j(v)},\max_{u\in\operatorname{pred}(v)}f_u\right\},\\
\rho_c&=\max_{r\in M_c}\rho_{v_{c,r}}.
\end{aligned}
\label{eq:readiness}
\end{equation}
A node without predecessors has $\rho_v=a_{j(v)}$. Start times satisfy $s_v\geq\rho_v$. A noncommunication node starts when its channel is free and completes at $f_v=s_v+\tau_v$. All participating collective nodes start together at $S_c\geq\rho_c$, so $s_{v_{c,r}}=S_c$. The completion time on rank $r$ is $f_{c,r}=f_{v_{c,r}}$. Completion releases that rank's downstream dependencies, making further computation or communication ready.

We use collective-level resource admission on the given physical interconnect. Starting operation $c$ reserves the communication channels of its participating endpoints and the directed links in $\mathcal L_c(\Pi)$. Each endpoint channel is released when its rank completes the collective; links remain reserved until all members complete, at $F_c=\max_{r\in M_c}f_{c,r}$. This execution mechanism admits operations sharing endpoints or paths according to resource availability. Transfers within an admitted collective follow the given collective algorithm.

Let $\mathcal U(t)$ denote endpoints occupied by communication and $\mathcal B(t)$ the reserved directed links. The set $\mathcal Q(t)$ contains collectives with $\rho_c\leq t$ that have not started. The admissible set is
\begin{equation}
\begin{aligned}
\mathcal F(t;\Pi)=\{c\in\mathcal Q(t):\;&
\mathcal P_c(\Pi)\cap\mathcal U(t)=\varnothing,\\
&\mathcal L_c(\Pi)\cap\mathcal B(t)=\varnothing\}.
\end{aligned}
\label{eq:admission}
\end{equation}
Equation~\eqref{eq:admission} checks both endpoint and link availability. Operations on shared paths constrain one another even when they belong to different jobs; operations using disjoint resources can run concurrently.

A communication scheduling plan $\theta$ comprises a priority rule and its parameters. Given $\Pi$, the rule uses features such as communication cost, execution dependencies, and job arrivals to assign priority $p_c(\Pi,\theta)$. When the admissible set is nonempty, the executor selects
\begin{equation}
c^*(t)\in\underset{c\in\mathcal F(t;\Pi)}{\arg\max}
\ p_c(\Pi,\theta).
\label{eq:priority}
\end{equation}
Ties follow a deterministic operation-ID order. After admission, the executor updates resource occupancy and continues selecting admissible operations. Computation and communication completions update dependency and resource states. Placement determines resource conflicts, while priorities determine progress through those conflicts; together, they affect subsequent readiness and job completion.

\subsection{Job Completion Time and Joint Planning}
\label{sec:objective}
Placement and priority choices propagate through execution dependencies and ultimately affect job completion. Mean JCT captures both communication duration and its consequences for subsequent computation.

Under plan $(\Pi,\theta)$, job $j$ finishes when its last execution node completes. JCT is measured from its original arrival:
\begin{equation}
\begin{aligned}
C_j(\Pi,\theta)&=\max_{v\in\mathcal V_j}f_v(\Pi,\theta),\\
\mathrm{JCT}_j(\Pi,\theta)&=C_j(\Pi,\theta)-a_j.
\end{aligned}
\label{eq:jct}
\end{equation}
This metric includes dependency waits, communication queues, and any startup delay introduced by the plan.

Let $\mathcal P$ contain mappings satisfying \eqref{eq:mapping}, and let $\Theta$ contain the priority rules and parameters supported by the executor. Denote the arithmetic mean of job JCTs by $\overline{\mathrm{JCT}}(\Pi,\theta)$. The joint planning objective is
\begin{equation}
\min_{\Pi\in\mathcal P,\ \theta\in\Theta}
\ \frac{1}{J}\sum_{j\in\mathcal J}\left[C_j(\Pi,\theta)-a_j\right],
\label{eq:objective}
\end{equation}
subject to the arrival, dependency, channel, and communication admission constraints above. The delivered $\theta$ specifies the executor's rule and parameters; the method and experimental setup describe the available rules. Complete execution determines $C_j$. Identical static traffic volumes can yield different completion times because of path overlap and progress ordering, so joint planning must assess placement--scheduling combinations over the entire job graph.

\section{Proposed OrbitWeaver Method}
\label{sec:method}
OrbitWeaver implements the joint objective in Section~\ref{sec:model} as an LLM agent process for candidate construction and simulation evaluation under a given inter-satellite interconnect, job graphs, and GPU allocations. Structural analysis exposes communication membership, path conflicts, and dependency progress to guide placement--priority combinations. Fast simulation supports exploration, while reference simulation checks complete execution and provides feedback for further revision. Planning ends with a joint plan validated by the reference simulator for the current instance.

\subsection{Overall Framework}
\label{sec:framework}
\begin{figure*}[!t]
\centering
\includegraphics[width=\textwidth]{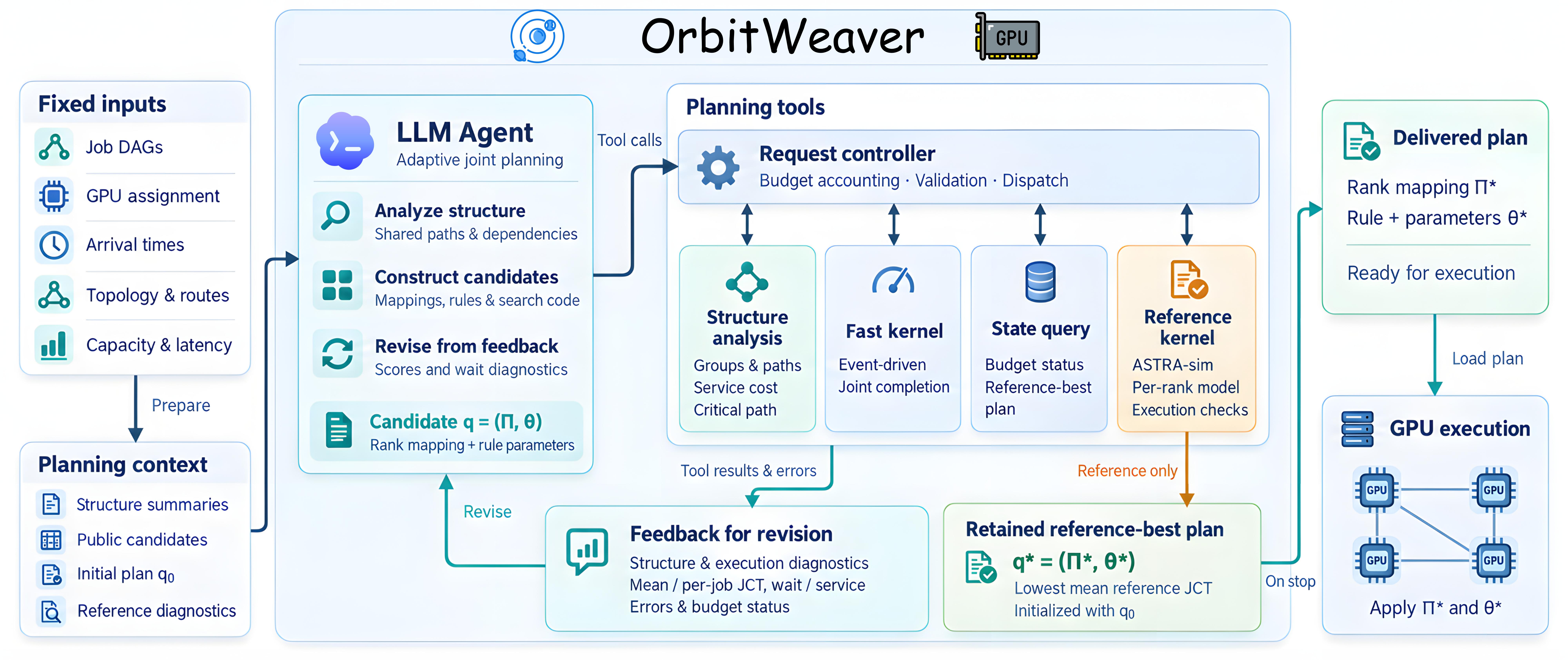}
\caption{OrbitWeaver's planning framework with simulation feedback.}
\label{fig:framework}
\end{figure*}

Placement and communication priorities jointly affect contention and dependency progress, so candidates must be evaluated over complete execution. In Fig.~\ref{fig:framework}, OrbitWeaver connects structural analysis, plan construction, and execution feedback through tool interfaces, allowing the agent to adapt its exploration to observed outcomes.

Each instance specifies job execution graphs, GPU allocations, arrivals, and interconnect parameters. Preparation extracts summaries of communication subgroups, path demand, and remaining critical paths, constructs a common set of placement candidates, and provides reference diagnostics for the initial placement. The initial plan $q_0=(\Pi_0,\theta_0)$ combines the given rank mapping with the Progress Alignment (PA) rule, providing a deliverable starting point.

The agent invokes three types of tools for joint plans $q=(\Pi,\theta)$. Structural analysis returns communication-group costs and path demand under specified mappings. Simulation executes candidates and returns mean and per-job JCTs with communication diagnostics. State queries expose the remaining request budget and the best reference-validated plan. A controller validates submissions, executes tool requests, and stores evaluation results.

All instances use the same task instructions and tool interfaces. The agent uses input summaries, previous candidates, simulation results, and failure feedback to choose what to analyze, how to revise candidates, and which evaluator to invoke. Tool order and candidate scope adapt to the current instance's feedback. The controller continually retains the feasible plan with the lowest reference-simulated mean JCT.

\subsection{Structural Analysis and Joint Candidate Construction}
\label{sec:candidate}
Execution feedback must inform concrete placement or scheduling changes to guide further exploration. OrbitWeaver connects path demand, execution cost, and queueing waits to help the agent identify ranks to remap and communication orders to reconsider.

Structural analysis first aggregates communication demand by job and participating ranks. For a given mapping, tools compute each subgroup's directed links and estimated contention-free service time, exposing shared paths, reduced-capacity links, and subgroups with concentrated communication costs. The initial plan's reference execution also provides proportional calibration of communication-group service estimates; the resulting factors are used in subsequent static estimates. The remaining critical-path estimate $K_c(\Pi)$ estimates the duration of the longest downstream path starting at collective $c$. It uses recorded durations for noncommunication nodes and mapping-dependent contention-free estimates for communication nodes.

Simulation diagnostics separate resource waiting from execution. Using the ready, start, and all-member completion times in Section~\ref{sec:model}, collective waiting and execution durations are
\begin{equation}
W_c=[S_c-\rho_c]_+,\qquad E_c=F_c-S_c,
\label{eq:waiting}
\end{equation}
where $[x]_+=\max(x,0)$. Tools aggregate $W_c$, $E_c$, and service estimates by job, membership, and reserved paths, and return per-job JCTs. Expensive subgroups indicate possible path changes, while high accumulated waiting suggests a need to examine progress ordering on shared resources. Together with remaining critical paths, these observations guide candidate revisions.

The agent can begin from a common candidate or the current best plan and modify rank permutations within jobs. Possible changes include regrouping communicating ranks on the physical network, changing how a logical communication ring is embedded in the assigned GPUs, or selecting different path-sharing relationships. It can also generate auxiliary search scripts to enumerate relevant permutations, merge structurally equivalent candidates, or explore local swaps and rotations. The current diagnostics determine the scope and screening features. Every candidate must satisfy the bijection constraint in \eqref{eq:mapping}.

Communication priority and placement form a candidate jointly. Fixed choices include PA, Shortest Processing Time (SPT), and ascending or descending job-ID priorities (Job-Asc and Job-Desc). The library also includes remaining-critical-path priority, critical-path priority normalized by the number of repeated trace windows, source-node ordering, and uniform priority. Parameterized progress and blend rules use
\begin{equation}
\begin{aligned}
p_c(\Pi;\alpha,\beta)
={}&(1-\beta)\bigl[K_c(\Pi)-K_{j(c)}^{\max}(\Pi)\bigr]\\
&-\beta\frac{i_c}{I_{\max}}\,\overline K(\Pi)-\alpha a_{j(c)}.
\end{aligned}
\label{eq:progress}
\end{equation}
Here, $K_j^{\max}(\Pi)$ is the largest remaining-critical-path estimate among job $j$'s collectives. The index $i_c$ is the communication node's ID in the original trace, and $I_{\max}=\max\{1,\max_c i_c\}$. If job $j$ contains $\nu_j$ repeated trace windows, then $\overline K(\Pi)=J^{-1}\sum_j K_j^{\max}(\Pi)/\nu_j$ scales the node-index term to time units. Parameters $\alpha$ and $\beta$ are finite real numbers; their signs and magnitudes determine the relative influence of arrivals, remaining paths, and source-node order.

For each mapping, the executor recomputes communication features and applies the selected rule and parameters. The agent can thus test a new placement with an existing priority or adjust priorities for new path-sharing relationships. Complete execution of the joint plan determines whether a change should be retained.

\subsection{Simulation Evaluation and Feedback-Guided Revision}
\label{sec:simulation}
Repeated candidate execution must balance evaluation cost with reliable completion-time assessment. OrbitWeaver uses an event-driven fast kernel for exploration and a reference kernel for execution validation, using both forms of feedback to guide further candidate selection.

The fast kernel preserves recorded noncommunication durations and dependencies and merges the participating nodes of each collective into a single joint node. Communication cost is computed from the given algorithm's synchronous rounds, routed byte loads, link capacities, and delays. An event executor advances dependencies and checks endpoint and link resources. A collective releases all participating resources and downstream dependencies through one completion event, reducing the cost of processing per-rank communication events.

The reference kernel extends ASTRA-sim's per-rank execution and native collective communication with the admission mechanism in Section~\ref{sec:admission}. It retains individual ranks' communication completion times, releasing each communication channel when its rank completes and reserved links when all members complete. Both kernels use identical mappings, priorities, recorded durations, arrivals, routes, and physical parameters. Reference execution additionally checks node completion, collective member pairing, and residual communication state to confirm complete execution.

The agent uses fast-kernel completion times, per-job outcomes, and diagnostics to select candidates for reference validation. Structural and execution evidence can also motivate a direct reference request to test a new path combination or priority choice. Reference JCTs and communication-group waits and durations update the retained plan and reveal whether estimated gains survive more detailed execution.

When reference performance worsens, the agent can select different subgroups, revise path combinations, or preserve a useful placement while changing priority. If the kernels rank candidates differently, reference diagnostics help reassess their potential. Execution feedback therefore informs both plan selection and the next round of structural analysis and candidate construction.

\subsection{Budget Control and Plan Delivery}
\label{sec:delivery}
Exploration must produce an executable result within finite planning time and simulation budgets. The controller manages evaluation requests and continually preserves a reference-validated plan for delivery when the session ends.

Let $B_f$ and $B_r$ be the fast and reference request limits. The controller deducts one request from the corresponding budget at submission; invalid plans, failed attempts, and cache hits all count against that budget. Structural analysis, local candidate construction, model interaction, and tool waiting contribute to planning time. Each submission is checked for mapping, rule, and parameter validity, and successful reference evaluations must also pass complete-execution checks.

Let $\mathcal V_k$ contain the successfully validated plans after processing the $k$th reference request, with $\mathcal V_0=\{q_0\}$. The controller retains
\begin{equation}
q_k^*\in\underset{q\in\mathcal V_k}{\arg\min}
\ \overline{\mathrm{JCT}}_{\mathrm{ref}}(q),
\label{eq:retained}
\end{equation}
where $\overline{\mathrm{JCT}}_{\mathrm{ref}}$ is the reference kernel's mean JCT. Successful reference requests extend $\mathcal V_k$; failed requests preserve error information for later revision. Retention consistently uses reference completion times, ensuring that the delivered placement--scheduling combination has completed evaluation.

Algorithm~\ref{alg:planning} summarizes offline planning. When the reference budget is exhausted or the session ends, the controller returns $q_k^*$. The executor loads its mapping and priority rule and advances jobs on the given interconnect using the readiness and admission conditions in Section~\ref{sec:admission}.

\begin{algorithm}[!t]
\caption{OrbitWeaver joint planning}
\label{alg:planning}
\renewcommand{\algorithmicrequire}{\textbf{Input:}}
\renewcommand{\algorithmicensure}{\textbf{Output:}}
\begin{algorithmic}[1]
\REQUIRE Jobs and interconnect; common placements; initial reference diagnostics and plan $q_0$; request limits $B_f,B_r$; session time limit.
\STATE Initialize context, remaining budgets, $\mathcal V_0=\{q_0\}$, and $q^*=q_0$.
\WHILE{the session is active and reference requests remain}
\STATE Agent selects an action using summaries, results, and remaining budgets.
\IF{the action is analysis or candidate construction}
\STATE Analyze or revise mappings, rules, or parameters; update context.
\ELSIF{the action is simulation evaluation}
\STATE Submit candidate and deduct the corresponding request; validate and invoke the selected kernel.
\STATE Add JCTs and diagnostics, or failure information, to context.
\IF{reference evaluation succeeds}
\STATE Add the candidate to $\mathcal V_k$ and update $q^*$ using \eqref{eq:retained}.
\ENDIF
\ELSE
\STATE Query budget or retained-plan state; update context.
\ENDIF
\STATE Agent revises its analysis targets and candidate direction from new feedback.
\ENDWHILE
\ENSURE Reference-validated plan $q^*=(\Pi^*,\theta^*)$ for loading and execution.
\end{algorithmic}
\end{algorithm}

\section{Evaluation}
\label{sec:evaluation}
\subsection{Experimental Setup}
\label{sec:setup}

\subsubsection{Simulation Environment and Workloads}
\label{sec:workloads}
The evaluation replays measured execution graphs over a parameterized optical interconnect in a discrete-event simulator. Computation retains recorded durations and dependencies, while communication is resimulated from collective membership, data sizes, algorithms, and physical routes. Following the separation of computation and network modeling in ASTRA-sim2.0~\cite{won2023astrasim2} and SimAI~\cite{wang2025simai}, this approach evaluates placement and communication order under different interconnect configurations using the same recorded workloads.

The orbital interconnect represents accelerator-equipped satellites in close formation, as envisioned by SunCatcher~\cite{aguera2025suncatcher}. Each node hosts one GPU endpoint and connects to two neighbors in a bidirectional ring, reflecting a limited optical-terminal budget; communication can therefore traverse shared multihop paths. Node spacing and nominal capacity use reference parameters from a public orbital data-center architecture artifact~\cite{orbitalAIDatacenterArtifact2026}. Topology, routes, and capacities remain fixed within each instance; power and cooling are assumed to sustain the recorded computation performance. Table~\ref{tab:interconnect} summarizes the interconnect configuration and parameter basis.

\begin{table*}[!t]
\centering
\caption{Orbital GPU interconnect configuration and parameter basis.}
\label{tab:interconnect}
\small
\renewcommand{\arraystretch}{1.22}
\setlength{\tabcolsep}{8pt}
\renewcommand{\tabularxcolumn}[1]{m{#1}}
\begin{tabularx}{\textwidth}{@{}>{\raggedright\arraybackslash}m{0.19\textwidth}>{\centering\arraybackslash}X>{\raggedright\arraybackslash}m{0.35\textwidth}@{}}
\toprule
\textbf{Parameter} & \textbf{Configuration} & \textbf{Basis}\\
\midrule
Cluster size & 16, 32, or 64 nodes; one GPU endpoint per node & Close-formation evaluation scales\\
\addlinespace[3pt]
Topology & Bidirectional ring with two neighbors per node & Limited optical-terminal budget\\
Neighbor separation $d$ & 300~m & Architecture artifact~\cite{orbitalAIDatacenterArtifact2026}\\
Single-hop latency & Approximately $2.001\,\mu\mathrm{s}$ & Propagation and terminal delay\\
\addlinespace[3pt]
Nominal capacity $B$ & 50~Gb/s per direction (6.25~GB/s) & Architecture artifact, converted per direction~\cite{orbitalAIDatacenterArtifact2026}\\
Capacity sweep & 50, 200, and 800~Gb/s per direction & Upper point: 800-Gb/s ground demonstration~\cite{aguera2025suncatcher}\\
\bottomrule
\end{tabularx}
\end{table*}

The workloads comprise the public H200 NeMo Mixtral-8x7B trace~\cite{chakraH200Artifact2026} and our A100 Transformer and ResNet-18 synchronous-training traces\footnote{The A100 traces will be made available at \url{https://github.com/luopeng69131/OrbitWeaver}.}, denoted \emph{H200 Mixtral}, \emph{A100 Transformer}, and \emph{A100 ResNet-18}. Each trace records eight ranks. The H200 trace captures communication subgroups of several sizes, while the A100 traces use randomly initialized models and synthetic inputs to capture workloads with different computation and communication demands. Table~\ref{tab:workloads} characterizes their original recording windows.

\begin{table}[!t]
\centering
\caption{Execution-graph, computation, and communication characteristics of the measured workloads.}
\label{tab:workloads}
\small
\renewcommand{\arraystretch}{1.16}
\setlength{\tabcolsep}{3pt}
\renewcommand{\tabularxcolumn}[1]{m{#1}}
\begin{tabularx}{\columnwidth}{@{}>{\raggedright\arraybackslash}m{0.33\columnwidth}*{3}{>{\centering\arraybackslash}X}@{}}
\toprule
\textbf{Feature} & \shortstack[c]{\textbf{H200}\\Mixtral} & \shortstack[c]{\textbf{A100}\\Transformer} & \shortstack[c]{\textbf{A100}\\ResNet-18}\\
\midrule
\multicolumn{4}{@{}l}{\textbf{Execution structure}}\\
Granularity & CPU/kernel & \mbox{Segment} & \mbox{Segment}\\
Window & \mbox{Full window} & 16 iterations & 16 iterations\\
Ranks & 8 & 8 & 8\\
Graph nodes & 2,093,878 & 6,408 & 1,032\\
\cmidrule(lr){1-4}
\multicolumn{4}{@{}l}{\textbf{Computation}}\\
\mbox{GPU compute nodes} & 265,424 & 4,352 & 768\\
\mbox{GPU time (GPU$\cdot$s)} & 19.608 & 4.741 & 1.697\\
\cmidrule(lr){1-4}
\multicolumn{4}{@{}l}{\textbf{Communication}}\\
Collectives & 12,640 & 256 & 32\\
Subgroups & 7 & 1 & 1\\
Subgroup size & 2, 4, 8 & 8 & 8\\
\mbox{Collective data (GiB)} & 375.253 & 5.070 & 0.697\\
\bottomrule
\end{tabularx}
\end{table}

Each configuration replays a single workload type, assigning eight GPUs to each job and preserving dependencies across one or two complete trace windows. Rank placement is restricted to each job's allocation. The scales use 2, 4, and 8 concurrent jobs, corresponding to 16, 32, and 64 GPU endpoints. Four operating conditions combine contiguous allocation with simultaneous arrivals, fragmented allocation with simultaneous arrivals, fragmented allocation with staggered arrivals, and fragmented allocation with staggered arrivals and reduced-capacity links. Combined with the three workloads, these yield 36 main configurations. Within each configuration, all methods use identical execution graphs, computation durations, GPU allocations, original arrivals, routes, and physical parameters.

\subsubsection{Baselines}
\label{sec:baselines}
The main performance comparison includes fixed communication rules, static rank placements, and published graph-mapping and communication-scheduling methods. Placement--rule combinations are named explicitly, such as Initial + PA.

\emph{Initial} uses a predefined deterministic mapping. Each job's eight assigned GPUs are sorted by physical ID; even-numbered jobs map ranks in ascending order and odd-numbered jobs in descending order. This provides a common starting placement for priority comparisons. \emph{Sum Placement} and \emph{Peak Placement} estimate link service demand by dividing routed communication bytes by link capacity. Both enumerate each job's eight-rank permutations. Sum Placement retains permutations tied for the minimum sum of link service demands, capped at 256; Peak Placement retains the best 64 permutations by maximum link service demand. Both combine per-job candidates using a beam of width 64 and select a joint placement by maximum link service demand. They are paired with PA and SPT, respectively.

\emph{Progress Alignment (PA)} is a model-generated heuristic frozen before the main experiments. For jobs satisfying its main communication-path compatibility conditions, it aligns progress using remaining critical paths and adjusts priorities using arrival spacing, source-node order, and path overlap. \emph{Shortest Processing Time (SPT)} prioritizes admissible collectives with shorter estimated contention-free service times. \emph{Job-Asc} and \emph{Job-Desc} prioritize lower and higher job IDs, respectively, breaking within-job ties by remaining critical-path length. Each of these four rules is evaluated with Initial placement.

\emph{Shortest Remaining Service First (SRSF)} follows Wang \emph{et al.}~\cite{wang2020communication} and their public implementation~\cite{hkbuSchedulingCode}. At communication admission, it prioritizes jobs with less remaining GPU computation, measured as the sum of unfinished GPU computation durations and updated when compute nodes complete. The rule uses fixed GPU allocations and the common endpoint and link admission mechanism.

\emph{SCOTCH}~\cite{pellegrini1996scotch} maps a source graph weighted by inter-rank communication bytes onto the physical ring, using inverse link capacities as distance weights in the target graph. We invoke SCOTCH 6.0.9 with quality and balance strategies once per job. Target terminals are restricted to that job's eight assigned GPUs, while other physical nodes remain available as routing relays. The resulting placement is evaluated with both PA and SRSF.

\emph{CASSINI}~\cite{rajasekaran2024cassini} selects rank placements using communication compatibility and coordinates jobs through launch-phase shifts. Isolated job runs generate periodic link-reservation profiles for placement selection and phase solving on the fixed-route ring. Equivalent demand constraints are merged, and phase constraints are solved jointly when the job-affinity graph contains cycles. The executor applies the selected placement and launch phases with PA. A100 profiles use recorded iteration boundaries; H200 profiles use the complete measured execution window, preserving each subgroup's demand on physical links. A single offline phase adjustment determines launch offsets under fixed GPU allocations, and any added delay contributes to JCT from the original arrival.

\emph{Crux}~\cite{cao2024crux} assigns communication priorities according to GPU computation intensity to improve GPU computation utilization. A job's intensity is estimated from the sum of recorded GPU computation times across its ranks divided by its bottleneck communication service demand. The latter is the maximum, over physical links, of routed bytes divided by link capacity. A correction for computation--communication overlap is obtained by comparing two-job periodic models under both priority orders. The corrected intensity determines a shared priority for all collectives of the job. With Initial placement and fixed routes, the executor selects among admissible collectives by priority and admits each collective without preemption.

\emph{Symphony}~\cite{binyahya2025symphony} coordinates collective execution across jobs by tracking request readiness and resource availability. With Initial placement, a collective enters a shared queue once all participating ranks are ready. The scheduler checks requests in arrival order and grants execution when the required endpoints and links are available. Blocked requests remain queued, while later requests using disjoint resources can proceed. An admitted collective executes without preemption, and its reserved links are released when all participants finish.

The following two searches are used for ablation and planning-mechanism analysis.

\emph{Candidate Search} evaluates combinations of the 16 common placement seeds and PA, SPT, Job-Asc, and Job-Desc, ranks them with fast simulation, and validates the leading candidates with reference simulation.

\emph{Local Search} starts from four placements and uses per-job feedback to explore neighborhoods comprising rank swaps, two-rank block swaps, cyclic shifts, and even/odd subgroup rotations.

\subsubsection{Implementation and Parameters}
\label{sec:implementation}
The implementation connects execution graphs, the interconnect model, and the planning controller so that methods evaluate candidates under common inputs and constraints. Graphs use Chakra format~\cite{chakraH200Artifact2026}. The reference kernel extends ASTRA-sim's per-rank execution and collectives~\cite{won2023astrasim2} with resource admission, using a congestion-aware analytical network backend and ring collective implementations. The C++ fast kernel advances events over merged collective nodes. Section~\ref{sec:simulation} describes their semantics; all final performance results use the reference kernel.

All simulations run as independent CPU processes on the same Linux server with Intel Xeon Platinum 8358P processors. The three planning methods share up to 12 concurrent tool slots and a cache that reuses completed evaluations for identical inputs and plans.

OrbitWeaver uses Codex CLI with model \texttt{gpt-6.1-sol} and reasoning effort \texttt{xhigh}. Each configuration provides 16 common placement seeds and initial reference diagnostics for PA, SPT, Job-Asc, and Job-Desc under Initial placement. The controller permits up to 64 additional fast requests and eight additional reference requests. Invalid attempts, failures, and cache hits consume requests. OrbitWeaver can select among ten priority rules and adjust their parameters. CASSINI's communication profiling and phase-solving costs, and Crux's profiling costs, are recorded separately.

OrbitWeaver's session time limits are 30, 60, and 120 min for 16, 32, and 64 GPUs, including tool waiting. Active tool requests may finish after the limit, with actual time recorded. Unless stated otherwise, OrbitWeaver variants use the same model, common candidates, initial diagnostics, and request limits. Reported plans come from the first independent session for each configuration. A timed-out session retains its best reference-validated plan; the common fallback is Initial + PA.

\subsubsection{Metrics and Aggregation}
\label{sec:metrics}
Evaluation separates the delivered plan's execution performance from the offline cost of obtaining it. Mean JCT is the primary performance metric; overhead experiments report planning and individual simulation times.

As defined in Section~\ref{sec:objective}, JCT is completion minus original arrival, including any plan-induced startup wait. The arithmetic mean over jobs in configuration $k$ under method $m$ is $\overline{\mathrm{JCT}}_{k,m}$. For workload $w$, let $\mathcal K_w$ contain its configurations. Its aggregate result is
\[
\overline{\mathrm{JCT}}_{w,m}
=\frac{1}{|\mathcal K_w|}
\sum_{k\in\mathcal K_w}\overline{\mathrm{JCT}}_{k,m}.
\]
Workload-level results weight configurations equally, with 12 per workload in the main comparison. Scale comparisons average the four conditions at each cluster size; condition comparisons average the three cluster sizes. Workloads are reported separately.

Planning time is the wall time from controller creation to plan delivery, including candidate construction, model interaction, tool calls, simulation waiting, failures, and timeout attempts. Common preparation is recorded separately. The overhead comparison uses the median across the same configuration set for each method.

Individual simulation time is the elapsed time from simulation process startup through completion-state checks. Fast and reference runs are paired by identical configuration and complete plan, deduplicated by actual process, and summarized by their respective median times.

\subsection{Overall Performance}
\label{sec:overall}
This experiment compares OrbitWeaver with fixed scheduling rules, static placements, and published scheduling methods to assess joint planning for concurrent GPU jobs on a limited-connectivity optical interconnect. All 13 methods cover the 36 configurations comprising three workloads, three scales, and four operating conditions. Table~\ref{tab:overall} reports aggregate mean JCT for each workload.

\begin{table}[!t]
\centering
\caption{Overall mean JCT across the three workloads.}
\label{tab:overall}
\small
\renewcommand{\arraystretch}{1.22}
\setlength{\tabcolsep}{2pt}
\renewcommand{\tabularxcolumn}[1]{m{#1}}
\begin{tabularx}{\columnwidth}{@{}>{\raggedright\arraybackslash}m{0.36\columnwidth}*{3}{>{\centering\arraybackslash}X}@{}}
\toprule
\shortstack[l]{Method\\(JCT/s)} & \shortstack[c]{H200\\Mixtral} & \shortstack[c]{A100\\Transformer} & \shortstack[c]{A100\\ResNet-18}\\
\midrule
\multicolumn{4}{@{}l}{\textbf{Scheduling Rules}}\\
Initial + PA & 561.353 & 14.108 & 2.019\\
Initial + SPT & 434.451 & 12.049 & 1.688\\
Initial + Job-Asc & 425.841 & 11.744 & 1.689\\
Initial + Job-Desc & 388.004 & 11.910 & 1.720\\
Initial + SRSF & 380.179 & 11.744 & 1.689\\
\cmidrule(lr){1-4}
\multicolumn{4}{@{}l}{\textbf{Static Placement Methods}}\\
Sum Placement + PA & 544.204 & 17.884 & 2.506\\
Peak Placement + SPT & 551.273 & 18.700 & 2.556\\
SCOTCH + PA & 896.812 & 49.783 & 6.299\\
SCOTCH + SRSF & 604.314 & 29.010 & 3.807\\
\cmidrule(lr){1-4}
\multicolumn{4}{@{}l}{\textbf{State-of-the-Art Methods}}\\
CASSINI & 519.542 & 14.479 & 2.006\\
Crux & 411.232 & 11.744 & 1.689\\
Symphony & 461.827 & 13.798 & 1.968\\
\midrule
\textbf{OrbitWeaver} & \textbf{345.390} & \textbf{11.016} & \textbf{1.585}\\
\bottomrule
\end{tabularx}
\end{table}

OrbitWeaver achieves the lowest aggregate mean JCT for H200 Mixtral, A100 Transformer, and A100 ResNet-18, at 345.390, 11.016, and 1.585~s, respectively. Across the three workloads, OrbitWeaver reduces aggregate mean JCT by 21.0--33.5\%, 6.2--16.0\%, and 19.5--25.2\% relative to CASSINI, Crux, and Symphony, respectively. Among fixed rules, Initial + SRSF performs best on H200 Mixtral at 380.179~s, still above OrbitWeaver. Static placement changes have mixed effects. Sum Placement + PA reduces JCT on H200 Mixtral but increases it on both A100 workloads relative to Initial + PA. Peak Placement + SPT and both SCOTCH combinations have higher JCT than Initial with the same priority rule for all three workloads. This comparison shows that OrbitWeaver translates joint rank placement and communication scheduling into lower aggregate completion times for concurrent jobs under fixed GPU allocations and limited optical connectivity.

\subsection{Performance Across Cluster Scales}
\label{sec:scale}
\begin{figure*}[!t]
\centering
\includegraphics[width=\textwidth]{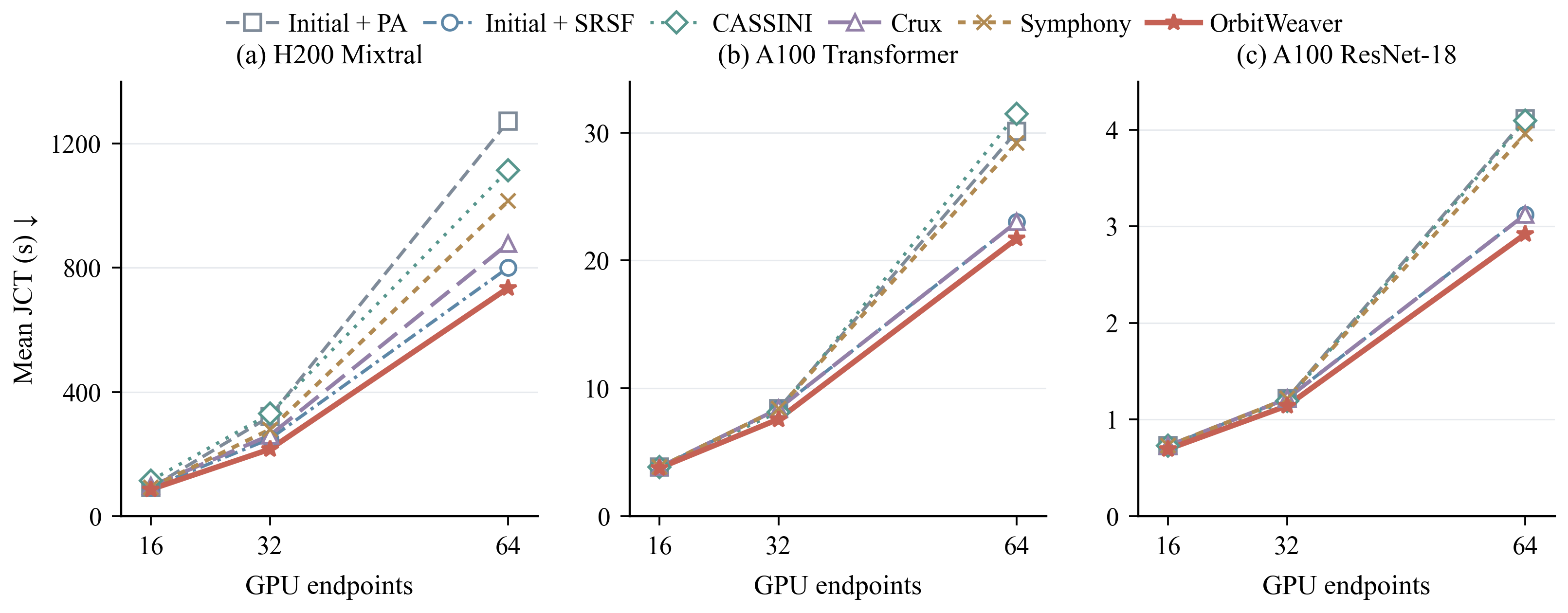}
\caption{Mean JCT across concurrent scales.}
\label{fig:scale}
\end{figure*}

This experiment assesses planning performance as the number of endpoints and concurrent jobs grows while each node retains two neighboring connections. The evaluated scales comprise 16, 32, and 64 GPU endpoints, corresponding to 2, 4, and 8 concurrent jobs. Results at each scale average the four operating conditions. Fig.~\ref{fig:scale} shows mean JCT separately for the three workloads.

Mean JCT rises with concurrent scale for all plotted methods and workloads. OrbitWeaver has the lowest grouped mean at every scale for all three workloads. On H200 Mixtral, Initial + PA, CASSINI, and Symphony exhibit steeper growth as scale increases. At 64 GPUs, Crux and SRSF achieve lower JCT than these three methods, while OrbitWeaver further reduces completion time. Differences between methods are small for the A100 workloads at 16 GPUs. At 64 GPUs, the gaps between OrbitWeaver and Initial + PA, CASSINI, and Symphony widen. Crux and SRSF produce similar grouped results but remain above OrbitWeaver. With the number of neighboring connections held fixed, OrbitWeaver consistently obtains lower mean JCT over the evaluated 16--64-GPU scales, supporting concurrent job execution on optical interconnects with limited connectivity.

\subsection{Performance Across Operating Conditions}
\label{sec:conditions}
\begin{figure*}[!t]
\centering
\includegraphics[width=\textwidth]{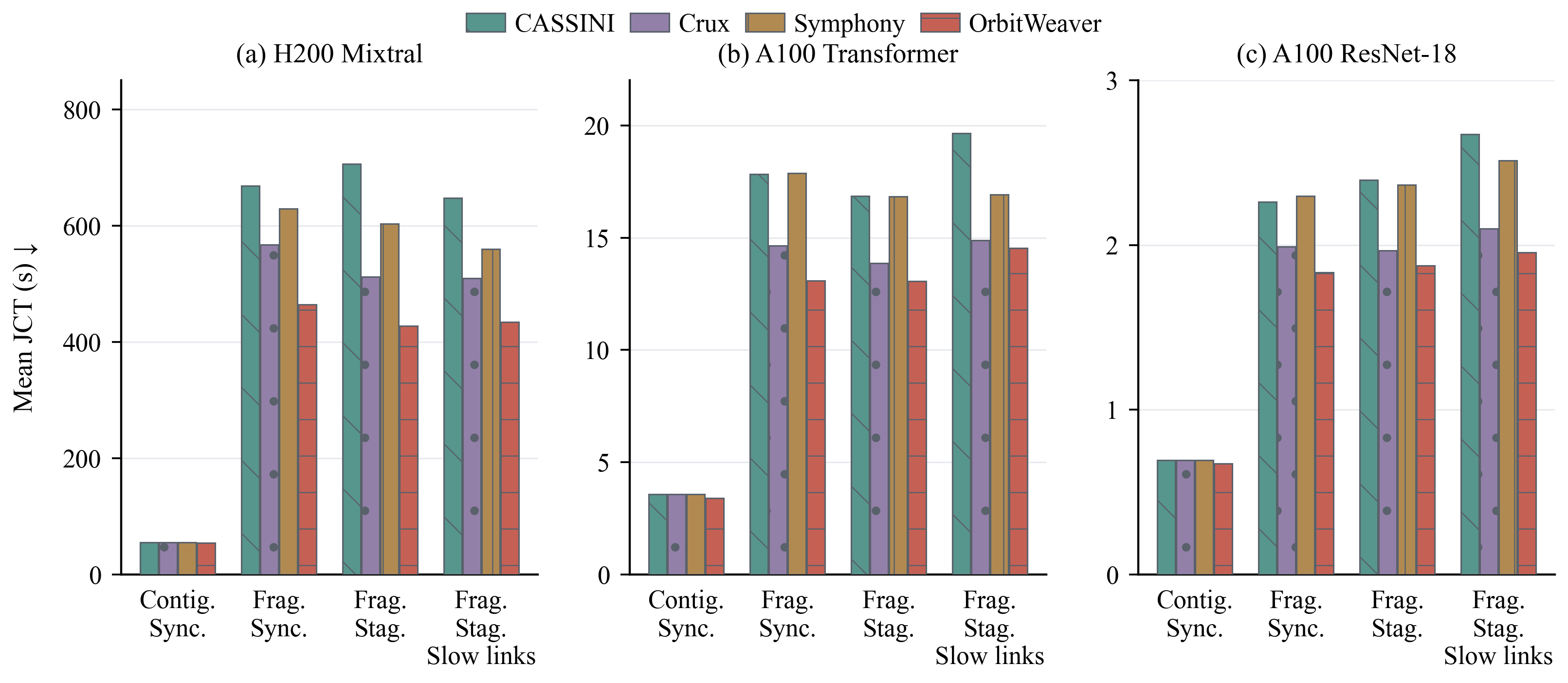}
\caption{Mean JCT across operating conditions.}
\label{fig:conditions}
\end{figure*}

This experiment examines the combined effects of GPU allocation, job arrivals, and static link capacities on planning performance. For each workload, results are grouped by operating condition and averaged over the three scales, comparing CASSINI, Crux, Symphony, and OrbitWeaver. In Fig.~\ref{fig:conditions}, Contig. and Frag. denote contiguous and fragmented GPU allocations, Sync. and Stag. denote simultaneous and staggered arrivals, and Slow links denotes static capacity reductions on specified directed links. Bars within each group follow the legend order.

Methods have similar mean completion times under contiguous allocation and simultaneous arrivals. Fragmented allocation increases the differences: with simultaneous arrivals, CASSINI and Symphony have higher mean JCT than Crux and OrbitWeaver, with a particularly large gap for H200 Mixtral. Under fragmented allocation with staggered arrivals, both with and without reduced-capacity links, OrbitWeaver retains lower grouped mean JCT than the other three methods. It also improves on Crux for both A100 workloads. These results show that OrbitWeaver can construct joint plans using endpoint allocations, arrival patterns, and link capacities to improve concurrent execution across the predefined optical-interconnect conditions.

\subsection{Link Bandwidth Sensitivity}
\label{sec:bandwidth}
\begin{figure}[!t]
\centering
\includegraphics[width=\columnwidth]{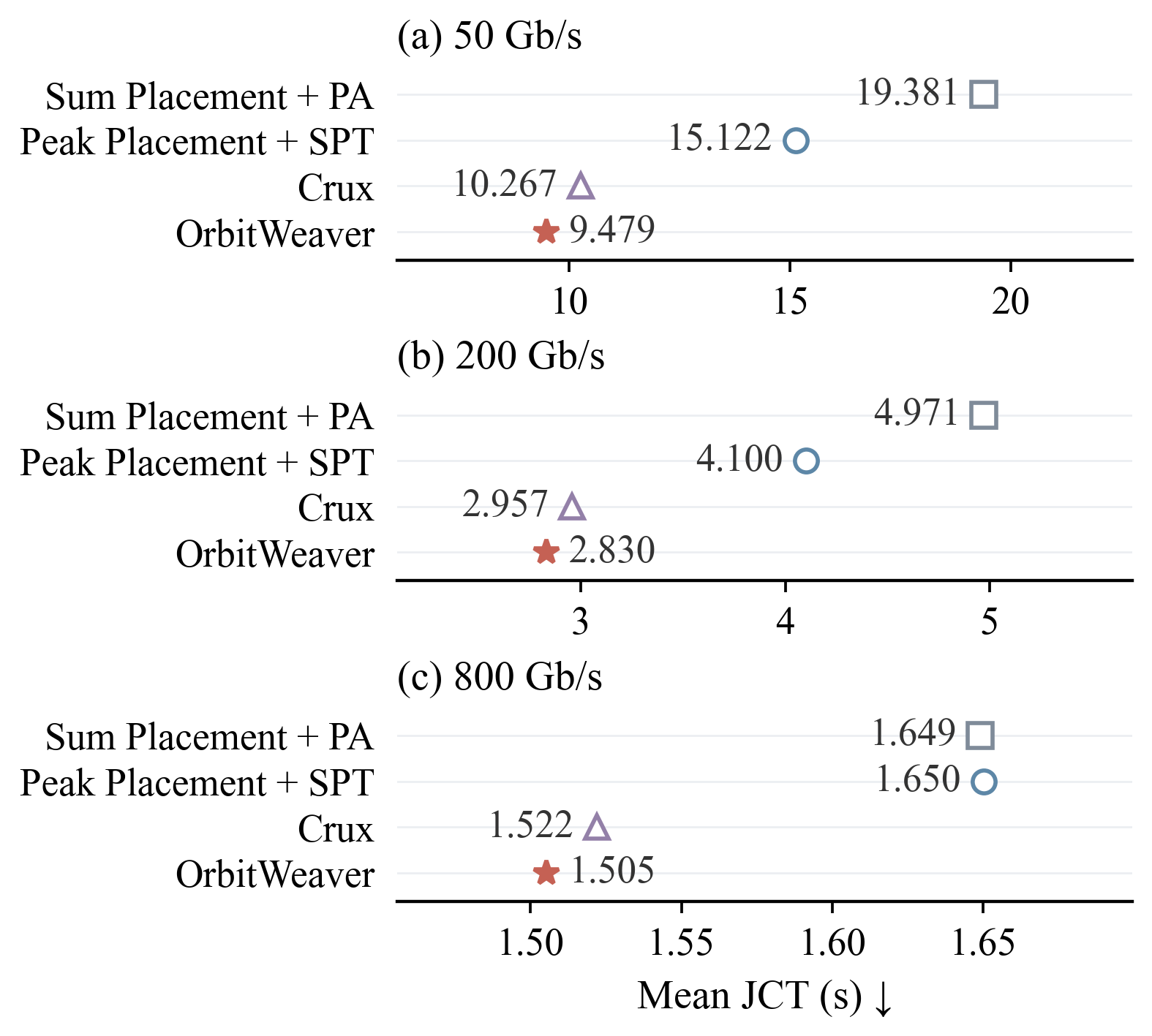}
\caption{Mean JCT at different inter-satellite link capacities.}
\label{fig:bandwidth}
\end{figure}

This experiment evaluates how inter-satellite optical-link capacity affects JCT and planning performance. It uses four A100 Transformer jobs on 32 GPUs with fragmented allocation, reduced-capacity links, and staggered arrivals. Nominal per-direction capacity is set to 50, 200, and 800~Gb/s, while GPU allocations, original arrivals, execution graphs, computation durations, routes, link latencies, and relative capacity reductions remain fixed. Fig.~\ref{fig:bandwidth} compares Sum Placement + PA, Peak Placement + SPT, Crux, and OrbitWeaver at each capacity.

Mean JCT decreases for all methods as bandwidth increases, and OrbitWeaver achieves the lowest value at all three capacities. Crux outperforms both static placements at every capacity, while OrbitWeaver further reduces completion time. At 50~Gb/s, Sum Placement + PA, Peak Placement + SPT, and Crux achieve 19.381, 15.122, and 10.267~s, respectively, compared with 9.479~s for OrbitWeaver. At 800~Gb/s, the corresponding results are 1.649, 1.650, 1.522, and 1.505~s. The absolute gaps between OrbitWeaver and each baseline narrow as capacity increases.
These results show that higher link capacity shortens execution, while OrbitWeaver provides larger performance gains at lower capacities. Joint planning is therefore particularly valuable when optical communication resources are constrained.

\subsection{Placement and Scheduling Ablation}
\label{sec:ablation}
This experiment examines the contributions of rank placement and communication scheduling to OrbitWeaver's performance. It fixes the 36 main configurations and plans from the first planning sessions. Starting from Initial + PA, the experiment substitutes OrbitWeaver's placement, its scheduling rule and parameters, or both. Placement-only plans retain PA, and scheduling-only plans retain Initial placement. Rule features are recomputed under the active mapping. Table~\ref{tab:substitution} compares mean JCT across the three workloads.

\begin{table}[!t]
\centering
\caption{Placement and scheduling ablation.}
\label{tab:substitution}
\small
\renewcommand{\arraystretch}{1.22}
\setlength{\tabcolsep}{2pt}
\renewcommand{\tabularxcolumn}[1]{m{#1}}
\begin{tabularx}{\columnwidth}{@{}>{\raggedright\arraybackslash}m{0.36\columnwidth}*{3}{>{\centering\arraybackslash}X}@{}}
\toprule
\shortstack[l]{Method\\(JCT/s)} & \shortstack[c]{H200\\Mixtral} & \shortstack[c]{A100\\Transformer} & \shortstack[c]{A100\\ResNet-18}\\
\midrule
Initial + PA & 561.353 & 14.108 & 2.019\\
Placement only & 471.985 & 14.169 & 1.960\\
Scheduling only & 385.574 & 11.723 & 1.704\\
\midrule
\textbf{OrbitWeaver} & \textbf{345.390} & \textbf{11.016} & \textbf{1.585}\\
\bottomrule
\end{tabularx}
\end{table}

Full OrbitWeaver achieves the lowest mean JCT for all three workloads. For H200 Mixtral, placement-only and scheduling-only substitution reduce JCT from 561.353~s to 471.985 and 385.574~s, respectively, while the joint plan further reduces it to 345.390~s. For A100 Transformer, placement alone yields 14.169~s, slightly above Initial + PA's 14.108~s, whereas the full plan reaches 11.016~s. Full OrbitWeaver also outperforms both separate substitutions on A100 ResNet-18. The results show that OrbitWeaver's joint planning of rank placement and communication scheduling achieves lower mean JCT for all three workloads.

\subsection{Planning with a Shared Rule Library}
\label{sec:shared-rules}
This experiment examines the value of OrbitWeaver's agent-guided planning and broader scheduling choices. Candidate Search and Local Search provide conventional references. The four-rule ablation retains OrbitWeaver's planning procedure but uses the searches' PA, SPT, Job-Asc, and Job-Desc rules, assessing plan construction under common choices. Comparison with full OrbitWeaver examines the complete rule library and parameter tuning. All methods share 12 configurations, initial information, and request limits. Table~\ref{tab:shared-rules} compares their mean JCT across the three workloads.

\begin{table}[!t]
\centering
\caption{Mean JCT with shared and full priority-rule libraries.}
\label{tab:shared-rules}
\small
\renewcommand{\arraystretch}{1.22}
\setlength{\tabcolsep}{2pt}
\renewcommand{\tabularxcolumn}[1]{m{#1}}
\begin{tabularx}{\columnwidth}{@{}>{\raggedright\arraybackslash}m{0.41\columnwidth}*{3}{>{\centering\arraybackslash}X}@{}}
\toprule
\shortstack[l]{Method\\(JCT/s)} & \shortstack[c]{H200\\Mixtral} & \shortstack[c]{A100\\Transformer} & \shortstack[c]{A100\\ResNet-18}\\
\midrule
Candidate Search & 353.837 & 12.247 & 1.763\\
Local Search & 340.024 & 13.478 & 1.762\\
\addlinespace[2pt]
\mbox{OrbitWeaver (four rules)} & 339.645 & 11.467 & \textbf{1.621}\\
\midrule
\textbf{OrbitWeaver} & \textbf{328.855} & \textbf{11.444} & \textbf{1.621}\\
\bottomrule
\end{tabularx}
\end{table}

OrbitWeaver outperforms both searches on all three workloads. On A100 Transformer, it achieves 11.444~s, compared with 12.247~s for Candidate Search and 13.478~s for Local Search. These results support the benefit of OrbitWeaver's agent-guided plan construction and refinement. Relative to the four-rule ablation, OrbitWeaver further reduces mean JCT from 339.645 to 328.855~s on H200 Mixtral and from 11.467 to 11.444~s on A100 Transformer. These comparisons support agent-guided joint planning and richer scheduling choices as complementary strengths of OrbitWeaver.

\subsection{Planning and Simulation Overhead}
\label{sec:overhead}
To relate execution gains to offline cost, we measure planning time over the 12 H200 Mixtral configurations for Candidate Search, Local Search, and OrbitWeaver, totaling 36 sessions. Individual kernel times use 227 paired fast and reference evaluations. Table~\ref{tab:overhead} reports the median times.

\begin{table}[!t]
\centering
\caption{Planning and simulation overhead for H200 Mixtral.}
\label{tab:overhead}
\small
\renewcommand{\arraystretch}{1.22}
\setlength{\tabcolsep}{3pt}
\renewcommand{\tabularxcolumn}[1]{m{#1}}
\begin{tabularx}{\columnwidth}{@{}>{\raggedright\arraybackslash}m{0.25\columnwidth}>{\raggedright\arraybackslash}X>{\centering\arraybackslash}m{0.28\columnwidth}@{}}
\toprule
Stage & Method/kernel & Median time\\
\midrule
\multicolumn{1}{@{}l}{\multirow{3}{*}{Planning}} & Candidate Search & 43.63~min\\
 & Local Search & 52.45~min\\
 & \textbf{OrbitWeaver} & 58.76~min\\
\midrule
\multicolumn{1}{@{}l}{\multirow{2}{*}{Simulation}} & Fast kernel & 5.405~s\\
 & Reference kernel & 323.865~s\\
\bottomrule
\end{tabularx}
\end{table}

OrbitWeaver's median planning time is 58.76 min, compared with 43.63 min for Candidate Search and 52.45 min for Local Search. Model analysis, candidate generation, and feedback interaction introduce additional planning work. The fast kernel's median individual runtime is 5.405 s, compared with 323.865 s for the reference kernel. Its lower cost supports repeated candidate evaluation. Together with the shared-rule planning results in Table~\ref{tab:shared-rules}, these results show that OrbitWeaver uses additional offline planning time to obtain better joint execution plans on the given interconnect, while the fast kernel reduces the cost of repeated exploration.

\begin{figure*}[!t]
\centering
\includegraphics[width=\textwidth]{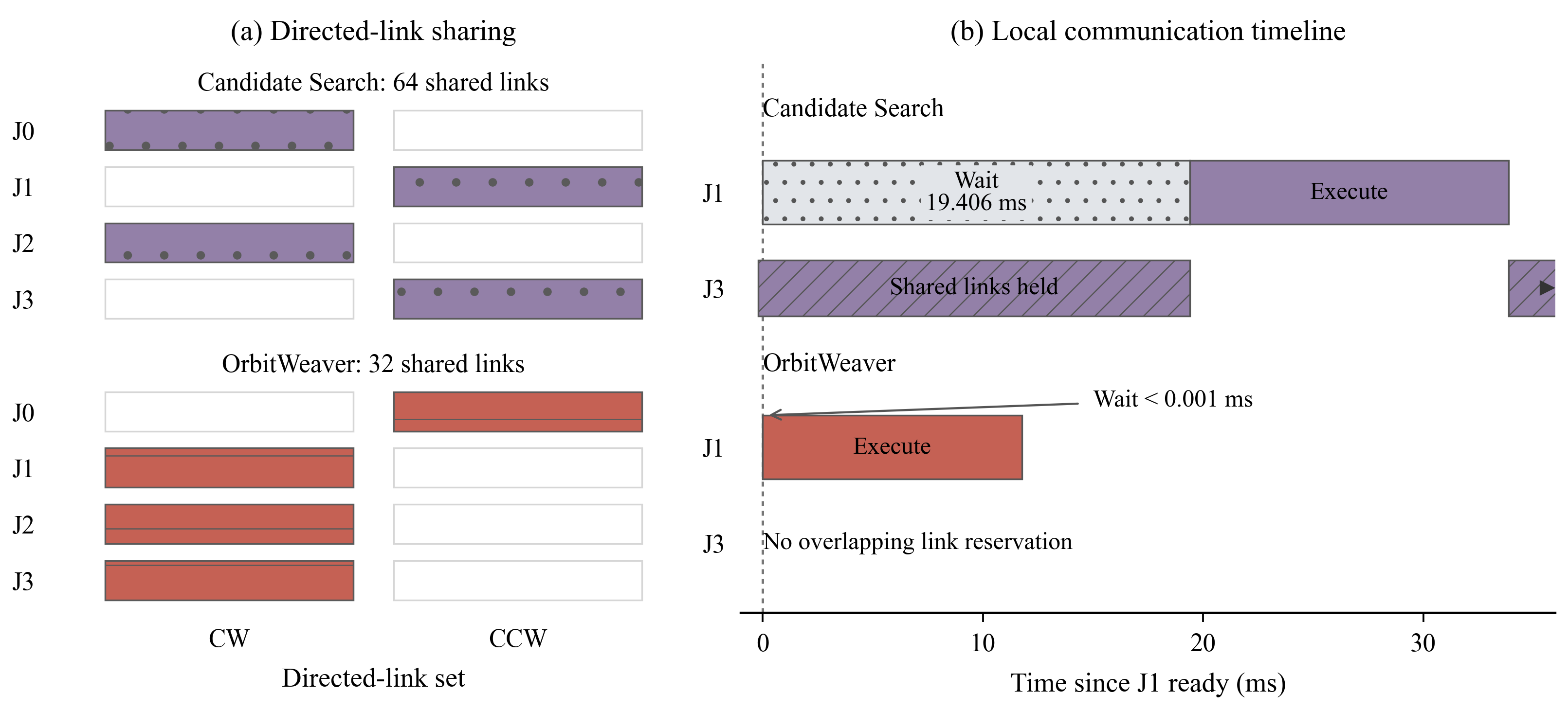}
\caption{Path sharing and communication blocking in the Transformer case.}
\label{fig:case}
\end{figure*}

\subsection{Case Study}
\label{sec:case}
This case study examines how OrbitWeaver coordinates rank placement and communication order to reduce shared-path contention and dependency blocking. It uses a 32-GPU, four-job A100 Transformer configuration with fragmented allocation, reduced-capacity links, and staggered arrivals. Table~\ref{tab:case} compares OrbitWeaver with Candidate Search and with plans replacing placement or scheduling alone.

Fig.~\ref{fig:case}(a) shows each job's reserved clockwise (CW) and counterclockwise (CCW) links over the execution window. Fig.~\ref{fig:case}(b) compares the two plans for the J1 collective with the largest waiting-time reduction.

\begin{table}[!t]
\centering
\caption{Plan comparison in the Transformer case.}
\label{tab:case}
\small
\renewcommand{\arraystretch}{1.22}
\setlength{\tabcolsep}{2pt}
\renewcommand{\tabularxcolumn}[1]{m{#1}}
\begin{tabularx}{\columnwidth}{@{}>{\raggedright\arraybackslash}m{0.30\columnwidth}>{\centering\arraybackslash}X>{\centering\arraybackslash}m{0.15\columnwidth}>{\centering\arraybackslash}m{0.18\columnwidth}@{}}
\toprule
Method & Placement & Rule & Mean JCT\\
\midrule
Candidate Search & Candidate Search & PA & 10.266\\
Placement only & OrbitWeaver & PA & 11.731\\
Scheduling only & Candidate Search & Progress & 10.266\\
\midrule
\textbf{OrbitWeaver} & OrbitWeaver & Progress & \textbf{9.479}\\
\bottomrule
\end{tabularx}
\end{table}

Replacing placement alone increases mean JCT from 10.266 to 11.731 s, while replacing scheduling alone leaves it at 10.266 s. Joint replacement gives OrbitWeaver 9.479 s, below the original and both separate substitutions.

In Fig.~\ref{fig:case}(a), J0 and J2 share CW links under Candidate Search, while J1 and J3 share CCW links. OrbitWeaver gives J0 exclusive use of CCW paths and places the other three jobs on CW paths, reducing cross-job shared directed links from 64 to 32.

In Fig.~\ref{fig:case}(b), the selected J1 collective waits 19.406 ms under Candidate Search. J3's reservation of the same link set covers almost its entire waiting interval. Under OrbitWeaver, J1 and J3 still share CW paths, but the collective starts after less than 0.001 ms of waiting, with no overlapping J3 reservation in the shown local window.

This case connects OrbitWeaver's plan to execution behavior: rank placement reorganizes path sharing across jobs, and the corresponding communication order reduces dependency blocking. Their coordination improves distributed GPU execution on the given optical interconnect.

\section{Conclusion}
\label{sec:conclusion}
We studied joint rank placement and communication scheduling for concurrent jobs in optically interconnected orbital GPU clusters. OrbitWeaver uses an LLM agent to combine execution-structure analysis with simulation feedback, producing joint plans that coordinate shared-link access and computation progress under fixed GPU allocations. Simulation experiments using H200 and A100 execution traces show reductions of 6.2--33.5\% in aggregate mean JCT relative to CASSINI, Crux, and Symphony. Ablations support the joint design and the complementary roles of agent-guided plan construction and richer scheduling choices. The bandwidth study further shows larger absolute JCT gains at lower optical-link capacities. These findings support OrbitWeaver as an effective approach to coordinating concurrent computation on orbital optical interconnects with limited connectivity. Future work will explore recursive self-improvement (RSI) to refine candidate-generation and planning procedures using validated feedback across tasks, aiming to improve adaptability to changing communication bottlenecks and reduce exploration cost.

\bibliographystyle{IEEEtran}
\bibliography{ref}
\onecolumn
\appendices
\section{Task Prompt and Tool Instructions for Codex Planning}
\label{app:prompts}
This appendix presents the initial task prompt and tool-interface instructions supplied to Codex for OrbitWeaver's offline planning sessions.

\subsection{Initial Task Prompt}
The initial instructions specify the planning objective, instance context, feasible decisions, and evaluation budget.

\begin{prompttemplate}{Initial planning prompt}
\footnotesize
\textbf{Task.} Jointly optimize rank placement and communication priorities for the given orbital GPU cluster instance. Minimize reference-simulated mean job completion time (JCT), measured from each job's arrival to its completion.

\smallskip\noindent\textbf{Context.} Read the supplied task interface and instance description: \emph{[GPU allocations; job arrivals; interconnect configuration; initial rank mapping; common placement candidates; initial reference results and communication diagnostics.]}

\smallskip\noindent\textbf{Constraints.} Permute ranks only within each job's eight assigned GPUs. Select a supported priority rule and finite parameters. Keep traces, dependencies, computing durations, arrivals, physical links, and the simulator fixed.

\smallskip\noindent\textbf{Budget.} Use at most \emph{[fast request limit]} fast evaluations and \emph{[reference request limit]} reference evaluations within \emph{[session time limit]}, including tool execution.

\smallskip\noindent\textbf{Planning.} Use structural analysis and simulation feedback to construct and revise joint candidates. Preserve successful plans and record unsuccessful hypotheses.

\smallskip\noindent\textbf{Return.} The best reference-validated placement and priority configuration, with a brief rationale for the planning decisions.
\end{prompttemplate}

The main experiments allow 64 fast and eight reference requests per instance. Session limits are 1,800, 3,600, and 7,200~s for 16, 32, and 64 GPU endpoints, respectively. Planning uses Codex CLI with model \texttt{gpt-6.1-sol} and reasoning effort \texttt{xhigh}.

\subsection{Tool and Feedback Instructions}
The following box summarizes the tool-interface instructions supplied at the start of each session. The initial task prompt and these instructions remain applicable throughout planning. Evaluation results return to the same conversation, where Codex selects subsequent analysis, candidate revisions, or validation requests. The controller does not send a separate continuation prompt after each evaluation.

\begin{prompttemplate}{Tool and feedback instructions}
\footnotesize
\textbf{Tools.} Use the supplied interface for structural analysis, fast and reference evaluations, remaining-budget queries, and best-plan retrieval.

\smallskip\noindent\textbf{Feedback fields.} \emph{[Mean and per-job JCTs; communication-group waiting and execution times; estimated service costs and shared paths; errors; current best reference-validated plan; remaining request budgets.]}

\smallskip\noindent\textbf{Revision.} Use structural analysis and execution feedback to select useful rank permutations and priority configurations. Keep effective priorities when revising placement. Retain successful plans and document hypotheses that did not improve reference JCT.

\smallskip\noindent\textbf{Validation.} Use the supplied tools for every simulation request. Fast scores can misrank candidates; reference-check promising changes before exhausting the session time. Failed attempts, invalid plans, and cache hits all consume the corresponding request budget.

\smallskip\noindent\textbf{Stopping and return.} Stop when no reference requests remain or the session ends. Retrieve the controller's best reference-validated plan and save the planning rationale. The retained plan includes the initial plan as a fallback.
\end{prompttemplate}

\end{document}